%% file: eprint.tex
\documentclass[12pt]{article}
\usepackage{graphicx}

\newcommand\pubnumber{---}
\newcommand\pubdate{\today}

\def\ucrfoot{\footnote{Physics \& Astronomy Department, 
University of California - Riverside, Riverside, CA, USA}}
\def\ucsbfoot{\footnote{Department of Physics, 
University of California - Santa Barbara, Santa Barbara, CA, USA}}
\def\ucsdfoot{\footnote{Department of Physics, 
University of California - San Diego, San Diego, CA, USA}}
\def\cubfoot{\footnote{Department of Physics, University of Colorado Boulder, Boulder, CO, USA}}

\def\Title#1{\begin{center} {\Large #1 } \end{center}}
\def\Author#1{\begin{center}{ \sc #1} \end{center}}

\newcommand\pubblock{\rightline{\begin{tabular}{l} \pubnumber\\
         \pubdate  \end{tabular}}}
\newenvironment{Abstract}{\begin{quotation}  }{\end{quotation}}
\newenvironment{Presented}{\begin{quotation} \begin{center} 
             PRESENTED AT\end{center}\bigskip 
      \begin{center}\begin{large}}{\end{large}\end{center} \end{quotation}}

\input econfmacros.tex
\newenvironment{Memoriam}{\begin{quotation} \begin{center} \begin{large}}{\end{large} \end{center} \end{quotation}}
\RequirePackage{ptdr-definitions}
\RequirePackage{heppennames2}
\usepackage[left]{lineno}
\begin{document}

\input{fourtop-definitions}
\begin{titlepage}
\pubblock

\vfill
\Title{Evidence for Four-Top Quark Production at the LHC}
\vfill
\begin{Memoriam}
To the memory of Stephen John Wimpenny, whose kindness and brilliance will be sorely missed.
\end{Memoriam}
\vfill
\Author{ Nicholas Manganelli \ucrfoot \cubfoot}
\Author{ Melissa Quinnan \ucsbfoot \ucsdfoot}
\Author{ on Behalf of the CMS Collaboration}
\vfill
\begin{Abstract}
The standard model production of four top quarks is predicted to have a cross section of the order of 12 fb. The CMS Collaboration presents new results on this rare production mechanism for Run 2 data collected in 2016 through 2018 at 13 TeV, considering event signatures containing zero to four electrons or muons. This is the first time the all-hadronic channel is investigated in the study of four top quarks, made possible through novel machine learning based data-driven background estimation techniques.
\end{Abstract}
\vfill
\begin{Presented}
$15^\mathrm{th}$ International Workshop on Top Quark Physics\\
Durham, UK, 4--9 September, 2022
\end{Presented}
\vfill
\end{titlepage}
\def\thefootnote{\fnsymbol{footnote}}
\setcounter{footnote}{0}

\section{Introduction}

The top quark, which is the heaviest fundamental particle in the standard model (SM),
was discovered in 1995 at the Tevatron. 
The simultaneous production of four top quarks in proton-proton
collisions is a very rare but expected process,
and previous analyses of collision data from 2015 to 2018 have approached statistical evidence~\cite{CMS:2019jsc,CMS:2019rvj} of its existence.
With a cross-section of 12\fb, the process is almost 5 orders of magnitude
rarer than \ttbar production, and serves as an important test of
high-mass QCD and high-multiplicty physics.
Additionally, enhancements to the production rate or kinematics of this process are
predicted by multiple beyond the standard model (BSM) theories, including
2 Higgs Doublet Models (2HDM) and extra-dimensions. The results presented
here combine new analyses of CMS~\cite{CMS:2008xjf} data collected in
2016 -- 2018 in the
all-hadronic (AH), semi-leptonic (SL), opposite-sign dilepton (OSDL) decay channels~\cite{CMS:2022uga} with results from the previously published same-sign dilepton and multilepton analysis~\cite{CMS:2019rvj}.

The three new analyses make orthogonal event selections based on the 
multiplicity of isolated electrons and muons (\ie leptons with minimal nearby 
tracks and energy deposits). All
channels select for a high multiplicity of anti-\kt radius 0.4 (AK4)
clustered jets, corresponding to a well-reconstructed \ttbar event
with several additional jets. An \HT cut of 500 -- 700\GeV significantly
reduces the main \ttbar background (and for AH, QCD multijet). Table~\ref{table:channels} summarizes each channel.

\begin{table}[!ht]
\begin{center}
\caption{Event selection and categorization criteria for events in each channel. Main background processes are estimated from data or Monte Carlo simulation (MC). The discriminating variable that is fit is either a Boosted Decision Tree (BDT) or \HT.}
\begin{tabular}{l|ccc}
 &  All-Hadronic & Semi-Leptonic &  OS Dilepton \\
\hline
\\[-1em]
Isolated Leptons &
Veto \Pe, \PGm &
1 \Pe or 1 \PGm &
$\Pe^\pm\!\Pe^\mp$, $\Pe^\pm\!\PGm^\mp$, $\PGm^\pm\!\PGm^\mp$ \\
\HT  &   $\geq 700$     &     $\geq 500$      &     $\geq 500$  \\
Number AK4 Jets &  9+     &     6+      &  4+ \\ 
Jet \pt Minimum & 35\GeV & 30\GeV & 30\GeV \\
Number \PQb-tags & 3+ & 3+ & 2+ \\
Additional Req. & $\geq 1 \PQt$-tag & $\pt^{miss} > 60\GeV$ & $m_{\it{ll}} > 20\GeV$ \\
Additional Req. &                     &                     & $|m_{\it{ll}} - 90\GeV| > 15\GeV$ \\
\hline
\\[-1em]
Major Bkg. & \ttbar + QCD & \ttbar & \ttbar \\
Main Bkg. Est. & Data-driven & MC & MC \\
Minor Bkg. Est. & MC & MC & MC \\
Categorization & \# \PQt-tags, & \# \PQb-tags, \# \PQt-tags, & \# \PQb-tags, \\
               & \HT          & \# jets, & \# jets, \\
               &              & lepton decay & lepton decay \\
\hline
Discriminating & & & \\
Variable & BDT & BDT & \HT \\
\hline
\end{tabular}
\label{table:channels}
\end{center}
\end{table}

\section{Single lepton analysis}
The SL analysis applies several techniques to separate signal and background
events and constraint background uncertainties. Events are categorized by
the year of data-taking, lepton decay channel (\Pe, \PGm), AK4 jet multiplicity
(6, 7, 8, 9, or $\geq$10), \PQb-tag multiplicity (3, $\geq$4), and \PQt-tag
multiplicity (0, $\geq$1). \PQt-tagging is accomplished with a Deep Neural
Net which uses kinematic, \PGb-tagging, and other information about triplets of
AK4 jets to determine the likelihood they come from a \PQt quark decay. With 3
hadronic \PQt decays per event, signal events have a higher chance of being
tagged. Simultaneously, \tttt signal events typically contain $\geq$9 jets,
and approximately 1 in 3 chance of all \PQb jets being tagged. Events with 2
\PQb-tags are used to derive corrections to simulation for higher \PQb-tagged
events.

The SL analysis uses a Boosted Decision Tree (BDT) trained to discriminate 
\tttt from \ttbar, with the fit performed on the shape of the BDT classifier.
More than 90\% of events in most categories are \ttbar: either \ttborbb (dominating the signal-enriched regions)
or \nonttborbb. The simulation cross-section for the former are scaled
based on dedicated CMS \ttbb measurements~\cite{CMS:2020grm}.
Amongst the most important input features to the BDT
are \PQb-tagging information from the highest \PQb-tagged jets, jet kinematic
information, \PQt-tag multiplicity and event shape information (e.g. \HT, centrality and Fox-Wolframm moments).

\begin{figure}[htb]
\centering
\includegraphics[width=0.32\textwidth]{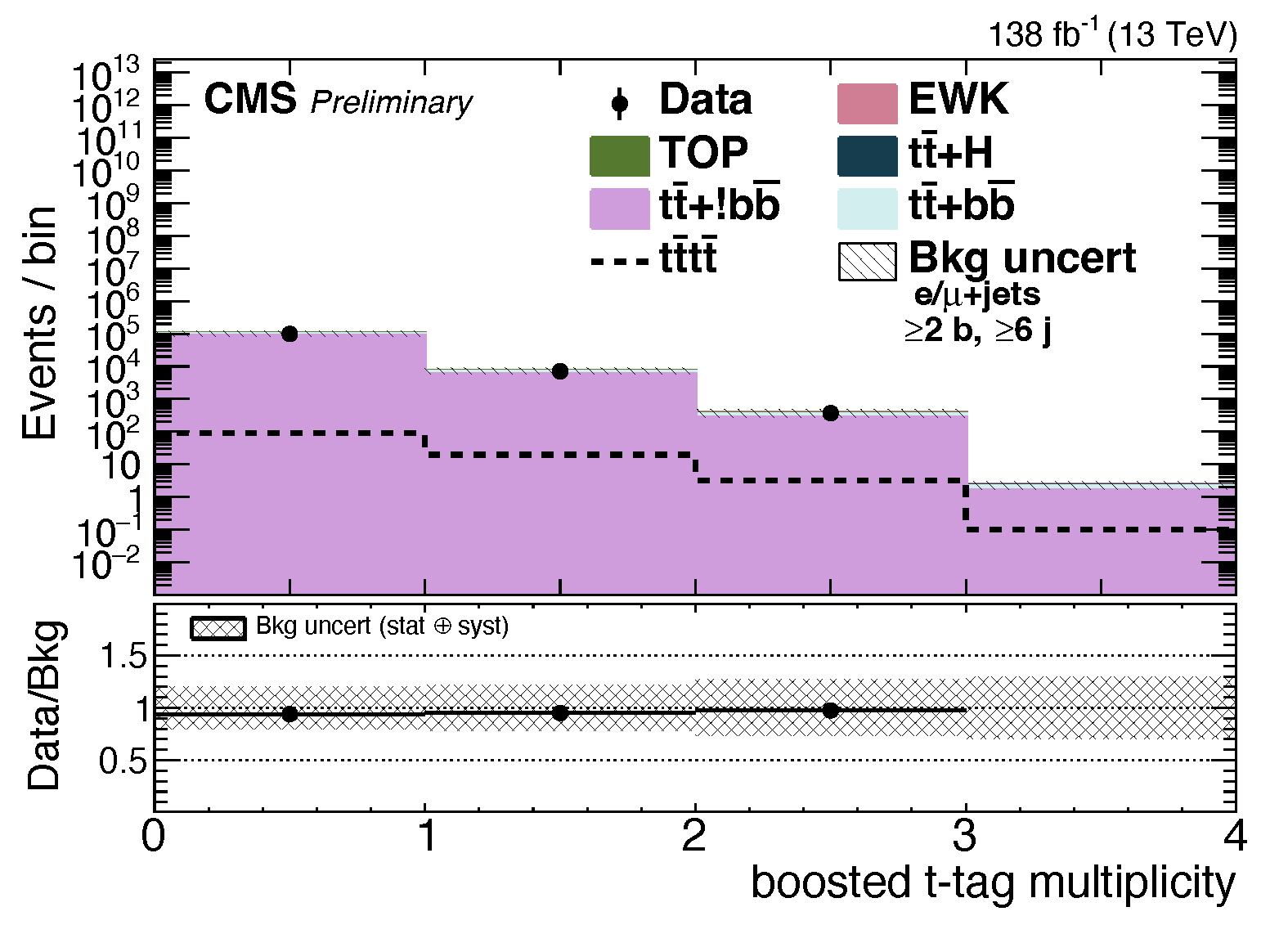}
\includegraphics[width=0.32\textwidth]{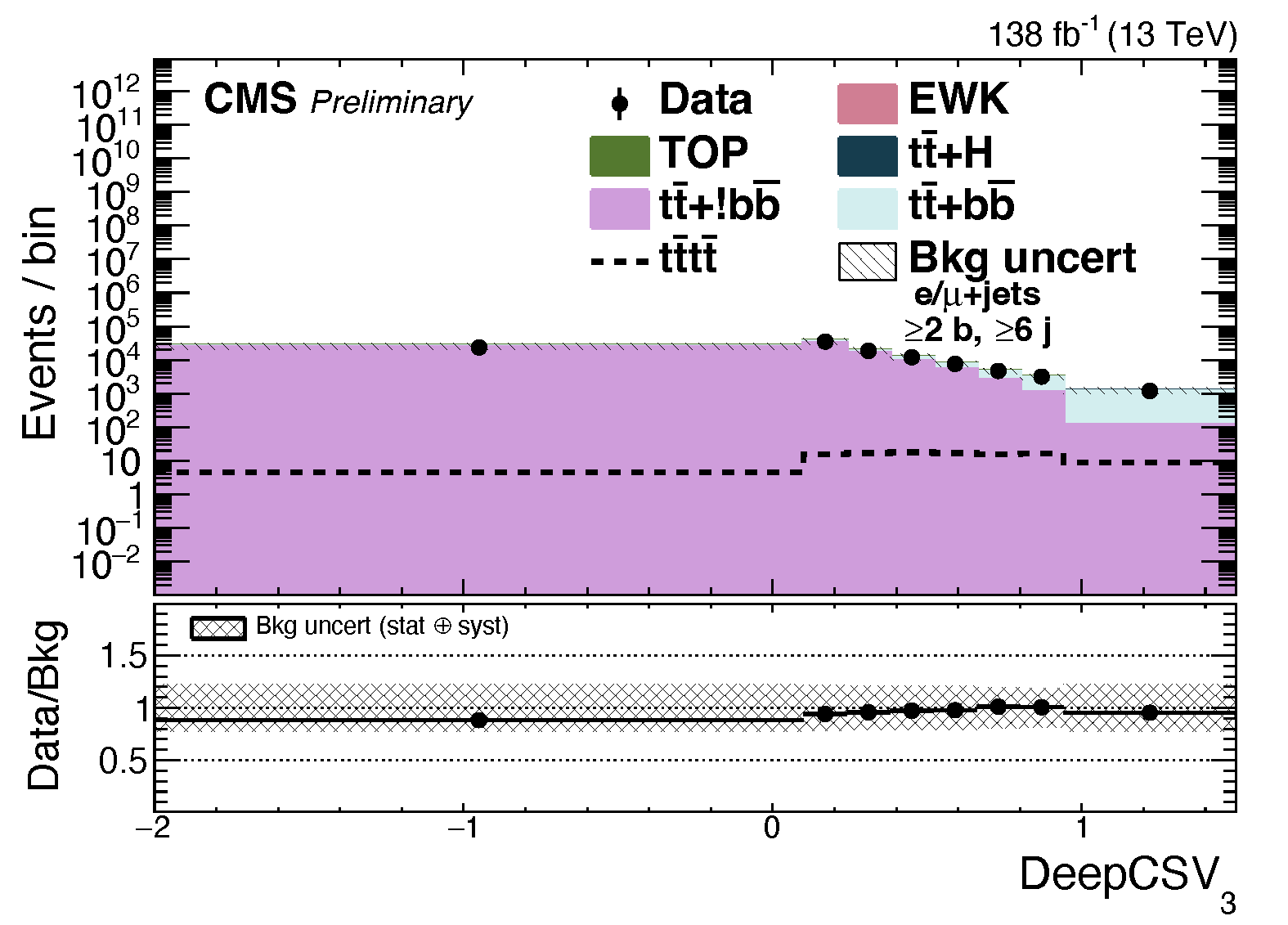}
\includegraphics[width=0.32\textwidth]{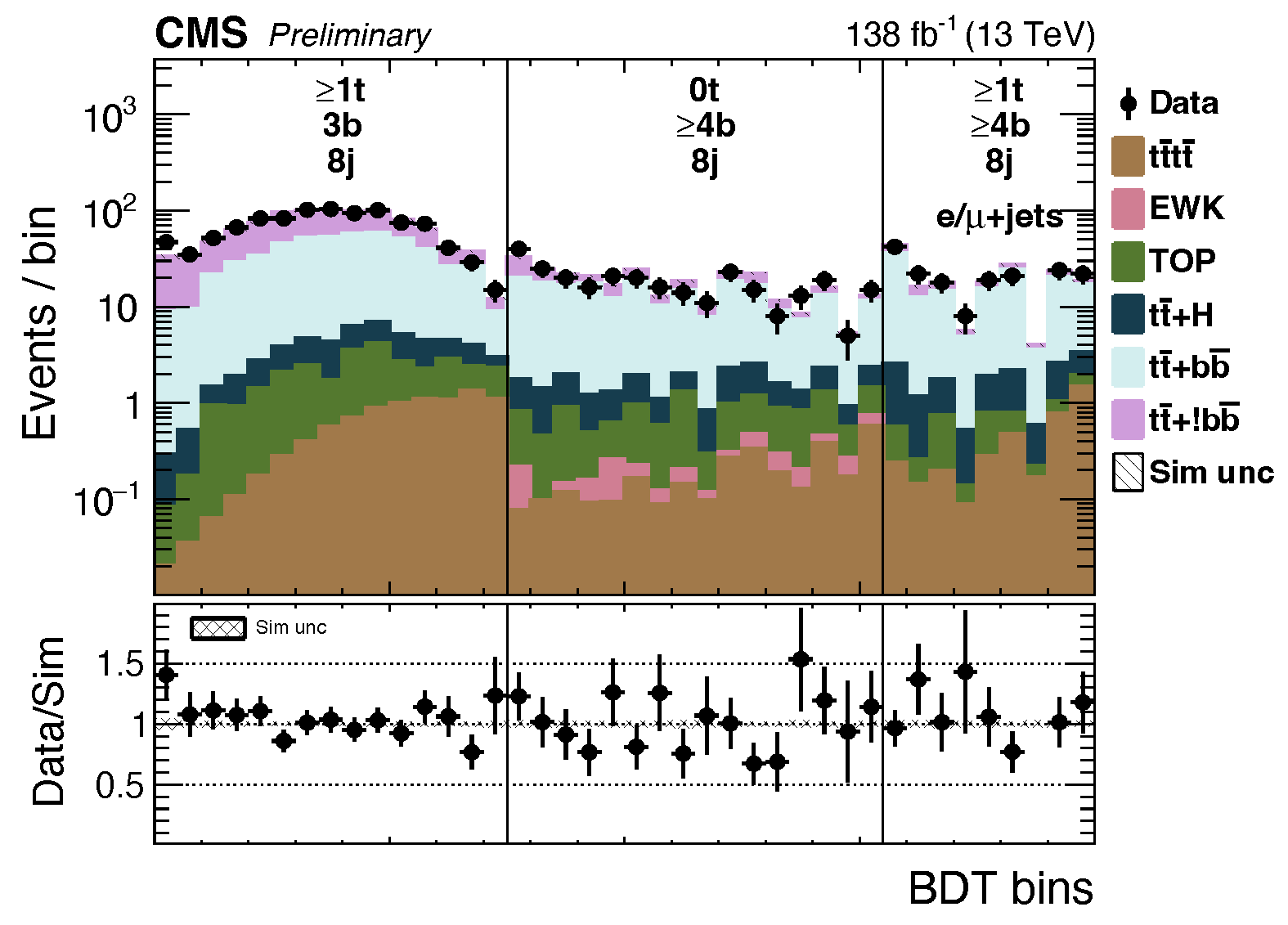}
\caption{Pre-fit distribution of \PQt-tag multiplicity (left) and third highest \PQb-tag discriminant (center), and post-fit distribution of 3 jet and $\PQt$-tag categories enriched in \tttt-signal (right) for the SL channel.}
\label{fig:SL_BDT_postfit}
\end{figure}

Shown in Fig.~\ref{fig:SL_BDT_postfit} are two pre-fit distributions for
the \PQt-tag multiplicity and third highest \PQb-tagged jet's discriminant,
and the post-fit distribution from 3 categories of events most enriched in 
\tttt-signal. These figures combine the two lepton channels and all 3 years
of data-taking. 

\section{Opposite-sign dilepton analysis}
The OSDL analysis strategy is similar to the SL channel. \HT is used
as the discriminating variable in the fit, and binned templates are generated
for events categorized by year, lepton decay channel, \PQb-tag multiplicity
(2, 3, $\geq$4) and jet multiplicity (4, 5, 6, 7, $\geq$8)
Akin to the SL analysis, the OSDL analysis contends primarily with \ttborbb
and \nonttborbb as backgrounds, with small contributions from \ttbar+\PH.

\begin{figure}[htb]
\centering
\includegraphics[width=0.32\textwidth]{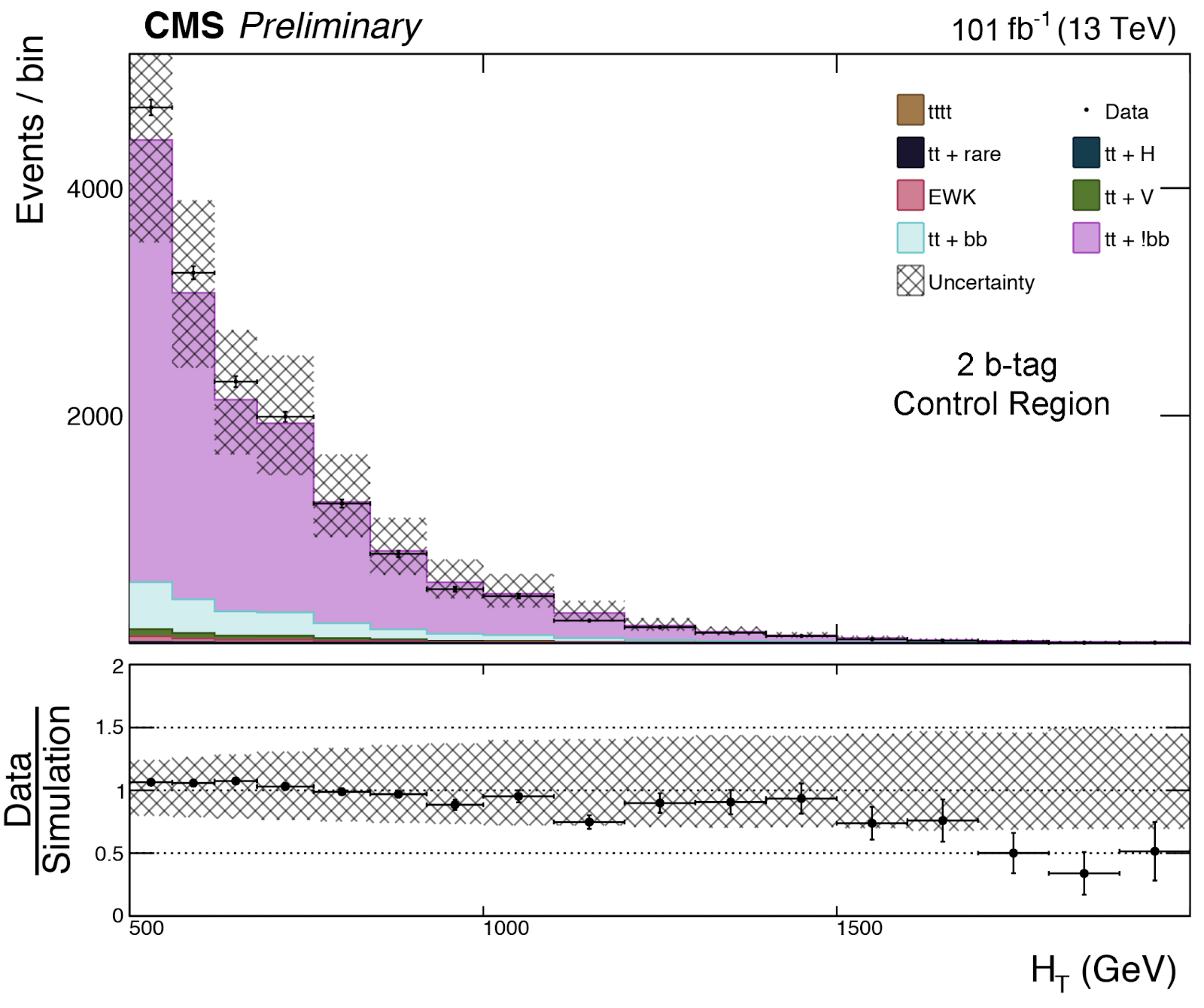} 
\includegraphics[width=0.32\textwidth]{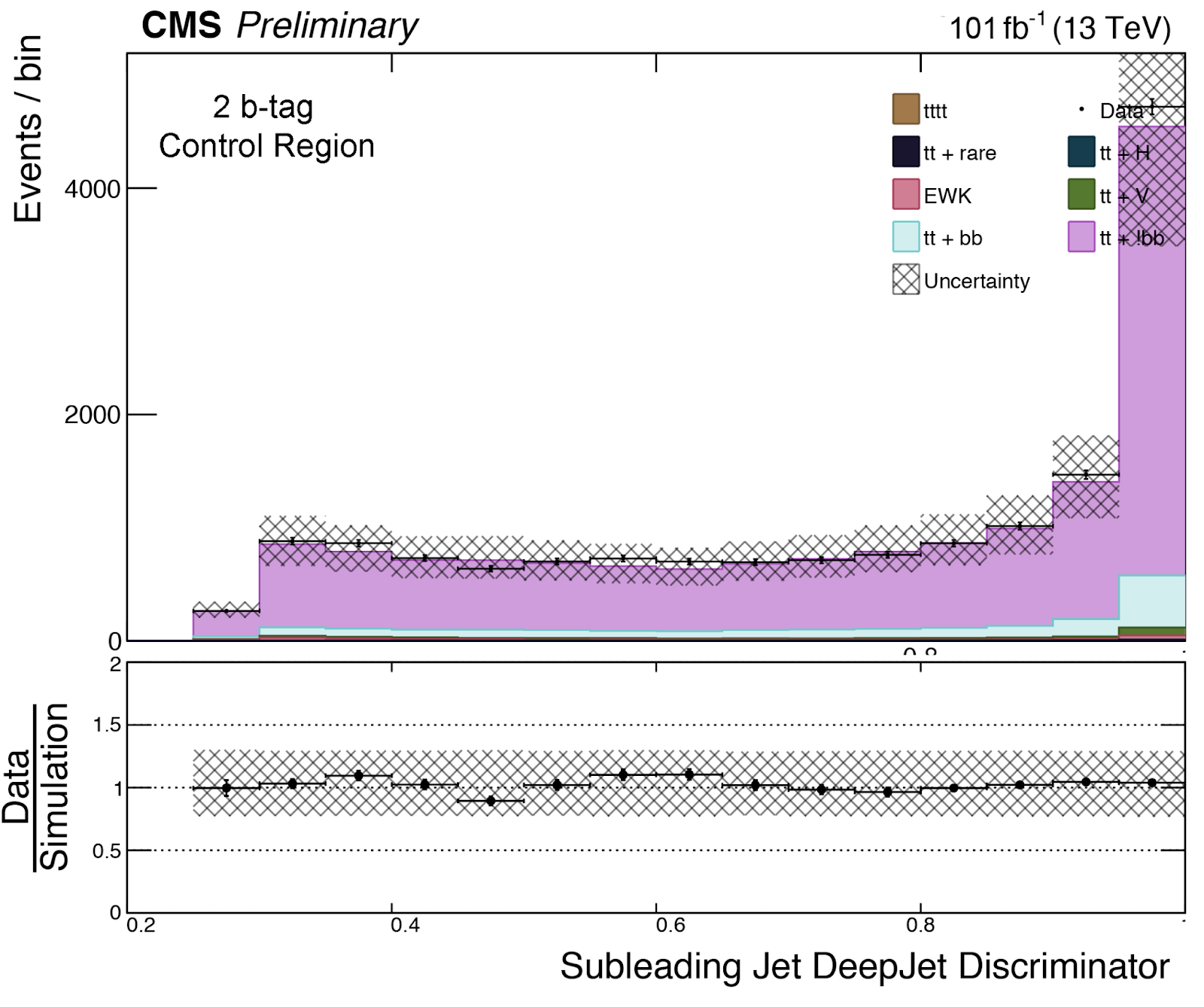} 
\includegraphics[width=0.3\textwidth]{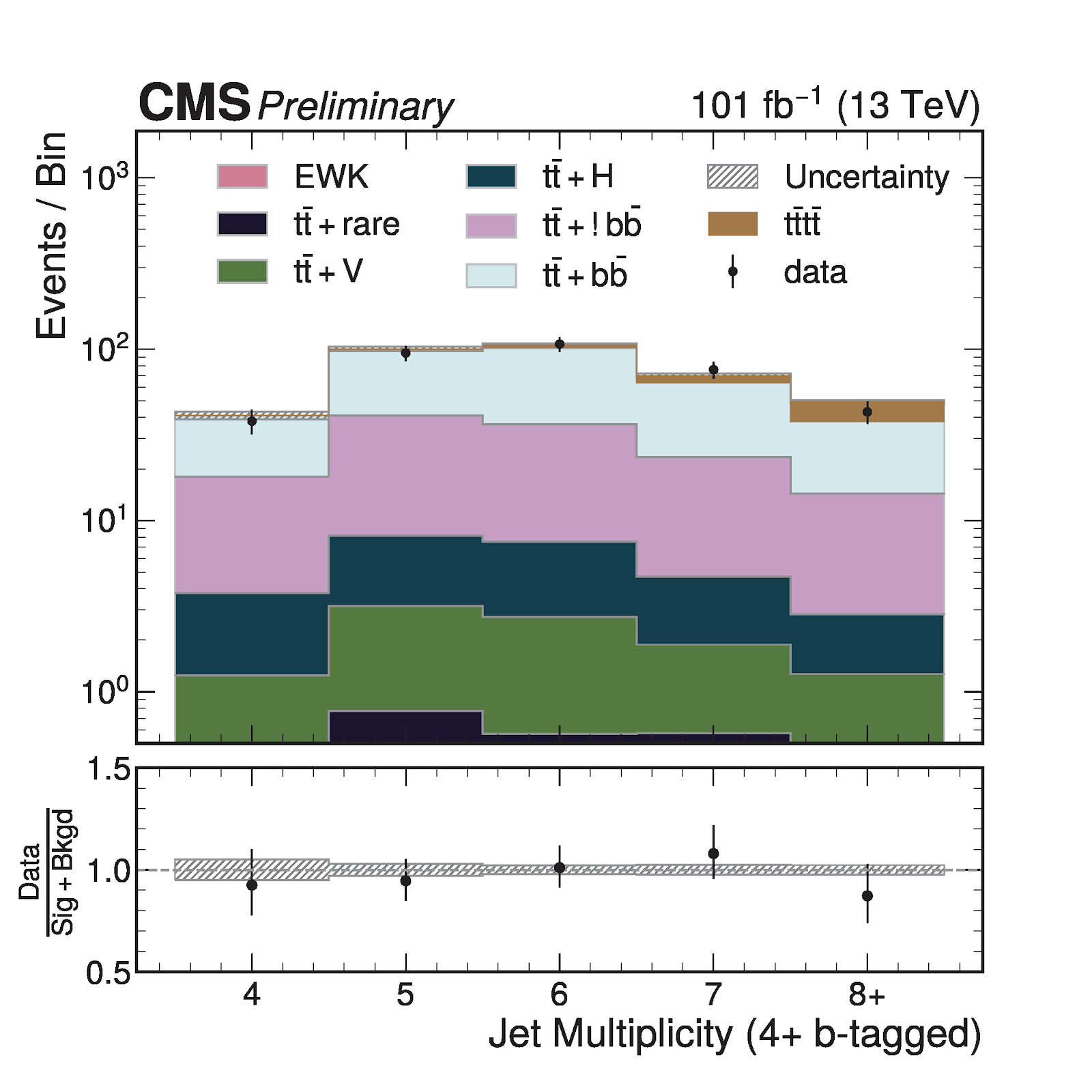} \\
\caption{
Pre-fit distributions of \HT (left),
the second highest \PQb-tagged jet discriminant (center),
and the post-fit category yields for $\geq$ 4 \PQb-tagged events (right)
in the OSDL channel.}
\label{fig:OSDL_pre_and_postfits}
\end{figure}

Pre-fit distributions of \HT, and the \PQb-tag discriminant of the second
highest tagged jet are shown in Fig.~\ref{fig:OSDL_pre_and_postfits}; modeling of variables is within systematic uncertainties.
The \HT distribution is softer in data than Monte Carlo, due to known
mis-modeling of the \PQt-quark \pt spectrum and 
initial / final state radiation (ISR/FSR), but is covered by shape
uncertainties in the fit.
The final post-fit distribution of category-yields for $\geq$4 \PQb-tags
is shown in the right of Fig.~\ref{fig:OSDL_pre_and_postfits}, 
combing 2017 and 2018 data for all 3 lepton decay channels.


\section{All-hadronic analysis}

\begin{figure}[htb]
\centering
\includegraphics[width=0.9\textwidth]{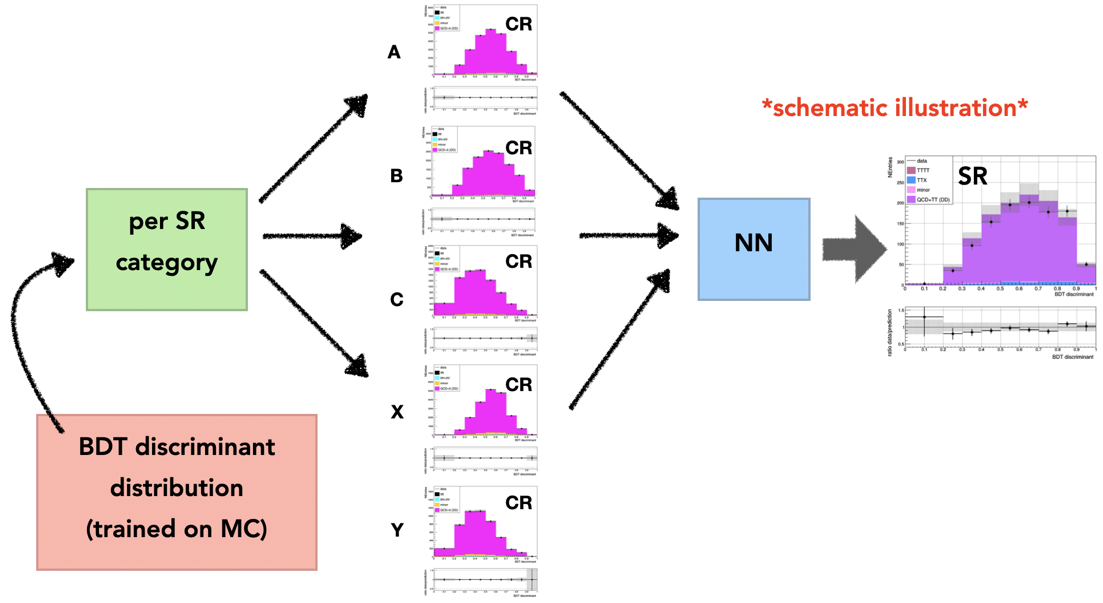} \\
\caption{Illustration summarizing all-hadronic data-driven \ttbar and QCD Multijet background estimation strategy.}
\label{fig:AH_ABCDnn}
\end{figure}

The AH channel employs a novel technique to estimate the background
which is principally \ttbar and QCD multijet.
As with the SL analysis, a  BDT is trained (on simulation) to separate \tttt from major backgrounds, which are hadronic \ttbar and QCD Multijet. Unlike the other channels, however, which use simulation to estimate contributions from major backgrounds, the all-hadronic channel uses a novel data-driven method based on a neural net (NN) to estimate the \ttbar and QCD Multijet backgrounds. Minor backgrounds are estimated from simulation. 

The data-driven background estimation is structured as illustrated in Fig.~\ref{fig:AH_ABCDnn}. 12 Signal Regions (SRs) defined by \PQt-tag multiplicity and \HT binning passing the requirements in Table~\ref{table:channels}. The strategy is to estimate the BDT discriminant distributions of the \ttbar and QCD multijet backgrounds from data, as indicated by the red box in the illustration. For each of these 12 SR categories, two tools are employed: one to estimate the shape of the BDT distribution for these backgrounds and one to estimate the number of expected background events, to which these BDT shapes are normalized. The illustrative histograms in Fig.~\ref{fig:AH_ABCDnn} represent the shape prediction. Five control regions (CRs) selected to be orthogonal to the SR by number of AK4 jets and \PQb-tags are defined. These CRs are defined by 2 or $\geq$3 tagged AK4 jets and 7,8,or $\geq$9 \PQb-tagged jets. A neural autoregressive flow
(ABCDnn~\cite{Choi:2020bnf}) is trained successively in each of these CRs, from least SR-like to most-SR like CR, to learn the transformation from a \ttbar MC input distribution to the \ttbar and QCD multijet target distribution. This target distribution is the data distribution in each CR with minor simulated contributions subtracted. The learned transformation is then applied to the \ttbar MC input distribution in the SR in order to predict the \ttbar and QCD multijet distribution in the SR. Training is done in each of the 3 top-tagged SR category per year; the BDT discriminant and \HT distributions are predicted simultaneously before splitting the BDT distributions into SR \HT categories to get the resultant \ttbar and QCD multijet predicted BDT distributions for each of the 12 SR top and \HT categories (this is done to ensure enough statistics in the training).

Because the number of events in the predicted target histogram produced by this method is the same number of events as was in the input histogram, this NN tool only predicts the shape of the backgrounds from data, not the normalization. A different tool is used to get the number of expected \ttbar and QCD multijet events per SR category. This uses the number of events in the same 5 CRs as previously described in a proportionality relation similar to the common "ABCD" method but with 5 CRs instead of 3. This "extended" ABCD method~\cite{Choi:ABCD} was found to give a better prediction of background yields than the traditional ABCD method. The predicted \ttbar and QCD multijet BDT discriminant histograms are then normalized to the number of predicted background events estimated by the extended ABCD relation. 

\begin{figure}[htb]
\centering
\includegraphics[width=0.9\textwidth]{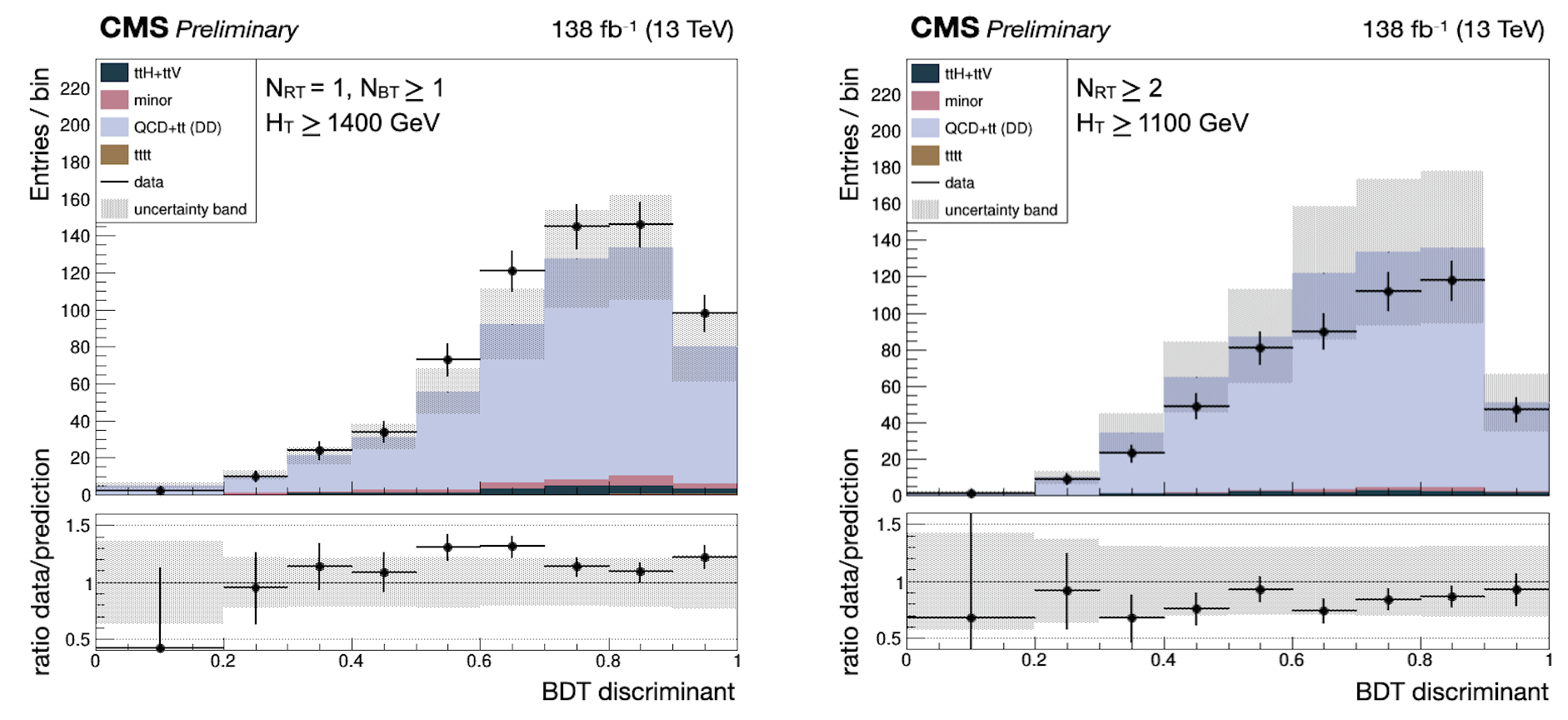}
\caption{VR distributions corresponding to the two most sensitive SR categories in the all hadronic final state, corresponding to high \HT and high tagged top multiplicity categories for all years 2016-2018 combined. Purple contributions come from the data-driven estimation of \ttbar and QCD multijet backgrounds. Discrepancies between data and background predictions are assigned as normalization and shape uncertainties to the corresponding SR categories.}
\label{fig:AH_BDT_VR}
\end{figure}

This background estimation strategy is validated in a validation region (VR) identical to the SR but with 8 tagged AK4 jets rather than 9 or more. The 12 VR categorizations are otherwise identical to the definitions in the SR. An example of two VR distributions can be found in Fig.~\ref{fig:AH_BDT_VR}. Any disagreements between the predicted BDT distributions and data in the VR categories are propagated as normalization and shape systematic uncertainties for this method.

\begin{figure}[htb]
\centering
\includegraphics[width=0.9\textwidth]{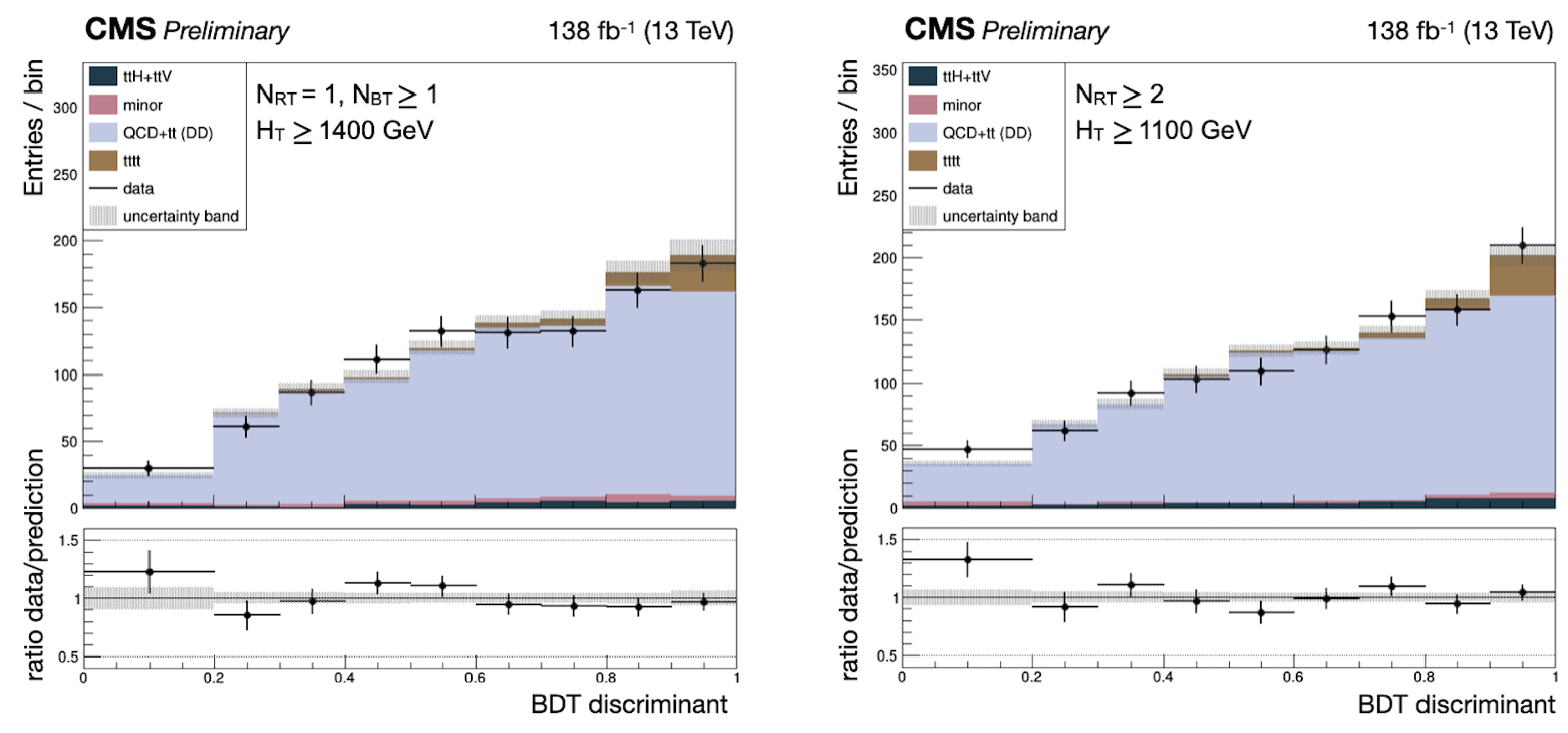}
\caption{Post-fit BDT distributions of the two most sensitive SR categories in the all hadronic final state, corresponding to high \HT and high tagged top multiplicity categories for all years 2016-2018 combined. Purple contributions come from the data-driven estimation of \ttbar and QCD multijet backgrounds.}
\label{fig:AH_BDT_postfit}
\end{figure}

Figure~\ref{fig:AH_BDT_postfit} shows the post-fit BDT discriminant distributions for the two most sensitive SR categories in the all-hadronic channel. Purple contributions come from the data-driven estimation of \ttbar and QCD multijet backgrounds, with shape and normalization estimated using the described NN and extended ABCD methods respectively. 




\section{Combination}

The three analyses described here are included in a simultaneous binned maximum
likelihood estimate (MLE) with the 2016 OSDL and 2016 -- 2018 Same-Sign Dilepton \& 
Multilepton (SSDL + ML); this is the first combination
of all \tttt decay channels (excluding hadronic $\tau$ decays).
The largest systematic uncertainties
include the rate of \ttborbb production, ISR/FSR, and the renormalization
and factorization scales (uncertainties due to limited order in the perturbative
expansion).
The combination of all channels attains an expected search significance of
3.2~s.d., with an observed significance of 3.9~s.d., The signal strength for the
combination as well as the AH, SL, OSDL, and SSDL/ML channels are
shown in Fig.~cite{fig:Combination}.
Consistent with the most recent ATLAS \tttt search~\cite{ATLAS:2021kqb},
a signal strength in excess of the SM expectation is observed.

\begin{figure}[htb]
\centering
\includegraphics[width=0.333\textwidth]{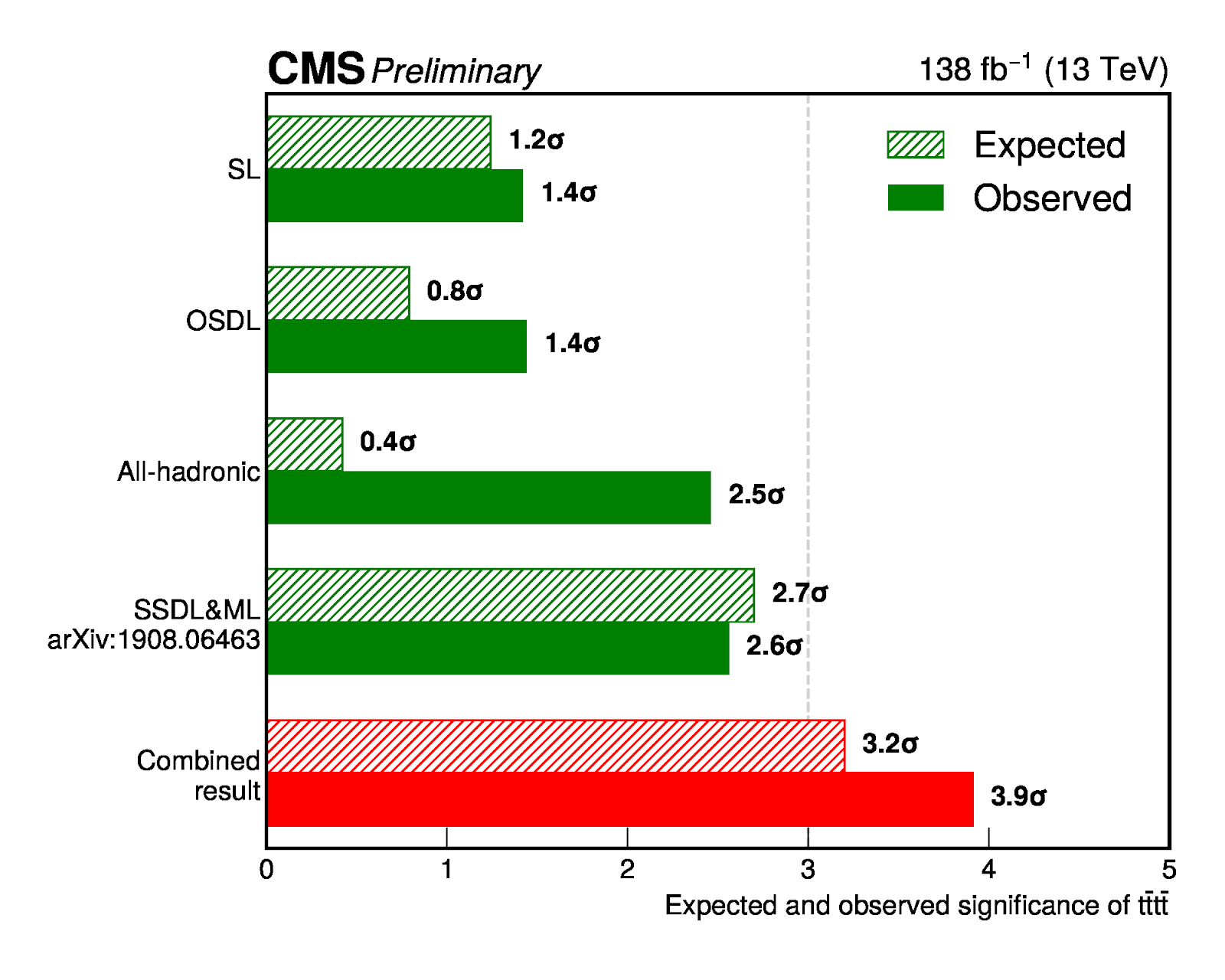}
\includegraphics[width=0.555\textwidth]{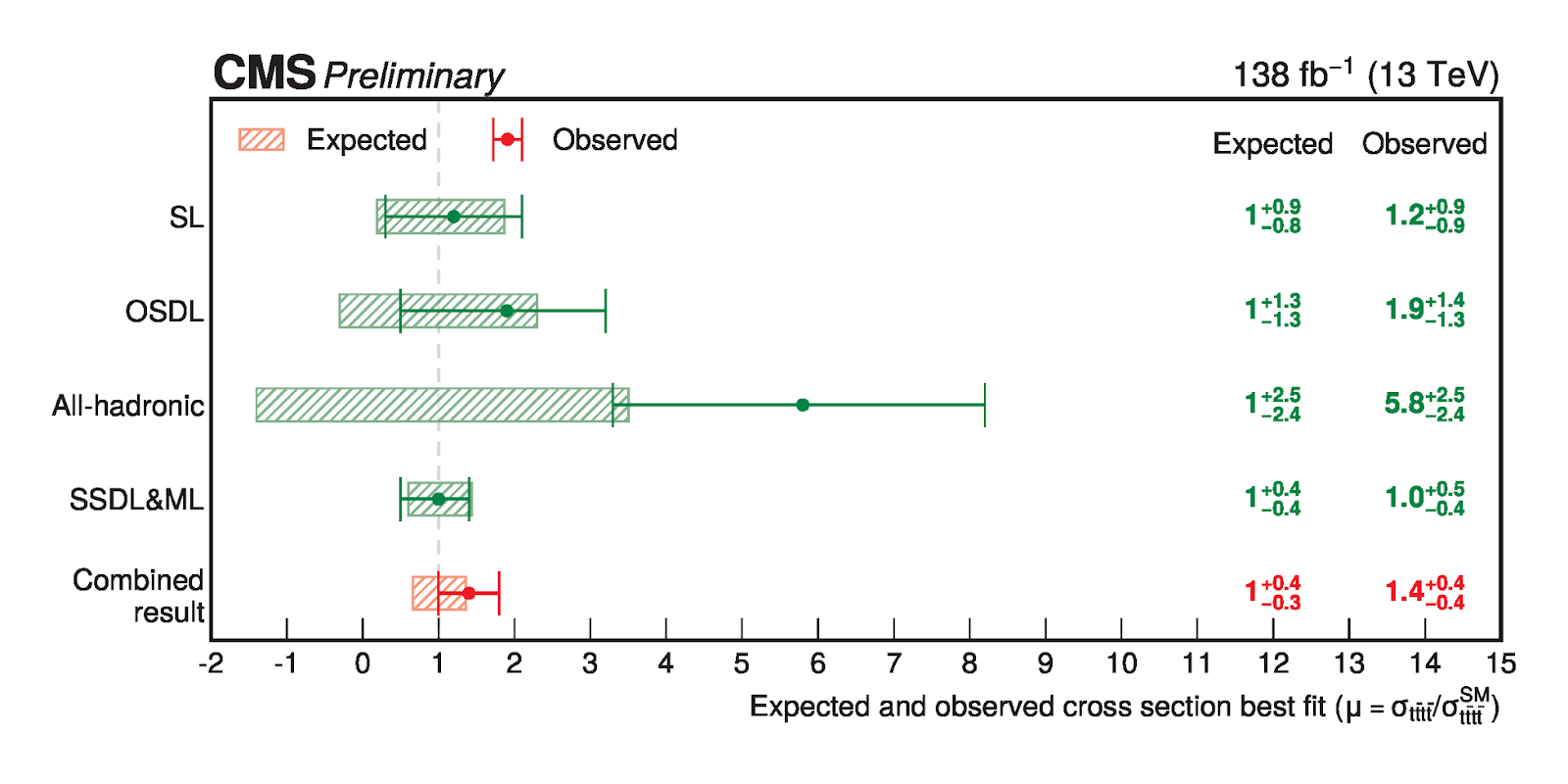}
\caption{Signal strength from the full combination, where each channel 
combines 2016 -- 2018 data. The observed significance for all decay
channels exceeds the SM expectation.}
\label{fig:Combination}
\end{figure}

\section{Summary}
This analysis combination is the first on \tttt to include all decay channels,
from 0 leptonically-decaying \PQt-quarks to 4, including the first successful
all-hadronic analysis. The CMS Collaboration achieves an expected and observed
search significance over 3~s.d., reaching experiment evidence of four-top production.



\end{document}

%% file: econfmacros.tex
\def\beq{\begin{equation}}
\def\eeq#1{\label{#1}\end{equation}}
\def\eeqn{\end{equation}}

\def\beqa{\begin{eqnarray}}
\def\eeqa#1{\label{#1}\end{eqnarray}}
\def\eeqan{\end{eqnarray}}

\let\bar=\overbar

\def\ie{{\it i.e.}}

\def\Dslash{\not{\hbox{\kern-4pt $D$}}}
\def\dslash{\not{\hbox{\kern-2pt $\del$}}}

\def\Pl{{\mbox{\scriptsize Pl}}}

\def\msb{{\bar{\ssstyle M \kern -1pt S}}}



%% file: fourtop-definitions.tex
\newcommand{\ttbb}{\ensuremath{\ttbar\bbbar}\xspace}
\newcommand{\ttb}{\ensuremath{\ttbar\PQb}\xspace}
\newcommand{\ttbbar}{\ensuremath{\ttbar\PAQb}\xspace}
\newcommand{\ttborbb}{\ensuremath{\ttbar\PQb}(\ensuremath{\PAQb})\xspace}
\newcommand{\nonttborbb}{non-\ensuremath{\ttbar\PQb}(\ensuremath{\PAQb})\xspace}
\newcommand{\ttcc}{\ensuremath{\ttbar\ccbar}\xspace}
\newcommand{\ttjj}{\ensuremath{\ttbar jj}\xspace}
\newcommand{\tttt}{\ensuremath{\ttbar\ttbar}\xspace}
\newcommand{\hdamp}{\ensuremath{h_\textrm{damp}}\xspace}
\newcommand{\apriorimu}{\ensuremath{2.9}}
\newcommand{\apriorimuup}{\ensuremath{+2.7}}
\newcommand{\apriorimudown}{\ensuremath{-1.4}}
\newcommand{\apriorixs}{\ensuremath{35}}
\newcommand{\apriorixsup}{\ensuremath{+32}}
\newcommand{\apriorixsdown}{\ensuremath{-16}}
\newcommand{\apriorisignif}{\ensuremath{0.64}}
\newcommand{\aposteriorimu}{\ensuremath{3.0}}
\newcommand{\aposteriorimuup}{\ensuremath{+2.9}}
\newcommand{\aposteriorimudown}{\ensuremath{-1.4}}
\newcommand{\aposteriorixs}{\ensuremath{36}}
\newcommand{\aposteriorixsup}{\ensuremath{+35}}
\newcommand{\aposteriorixsdown}{\ensuremath{-17}}
\newcommand{\aposteriorisignif}{\ensuremath{0.69}}
\newcommand{\observedmu}{\ensuremath{6.2}}
\newcommand{\observedxs}{\ensuremath{74}}
\newcommand{\observedsignif}{\ensuremath{1.9}}
\newcommand{\lhcobservedmu}{\ensuremath{X.XX}}
\newcommand{\lhcobservedxs}{\ensuremath{X.XX}}
\newcommand{\lhcobservedsignif}{\ensuremath{X.X}}
\newcommand{\njets}{N\ensuremath{_{\textrm{j}}}\xspace}
\newcommand{\njetsw}{N\ensuremath{_{\textrm{j}}^{\textrm{W}}}\xspace}
\newcommand{\nltags}{N\ensuremath{_{\textrm{tags}}^{\textrm{L}}}\xspace}
\newcommand{\nmtags}{N\ensuremath{_{\textrm{tags}}^{\textrm{M}}}\xspace}
\newcommand{\nttags}{N\ensuremath{_{\textrm{tags}}^{\textrm{T}}}\xspace}
\newcommand{\htb}{H\ensuremath{_{\textrm{T}}^{\textrm{b}}}\xspace}
\newcommand{\htrat}{H\ensuremath{_{\textrm{T}}^{\textrm{Rat}}}\xspace}
\newcommand{\httwom}{H\ensuremath{_{\textrm{T}}^{\textrm{2M}}}\xspace}
\newcommand{\thirdjetpt}{p\ensuremath{_{\textrm{T}}^{\textrm{Jet3}}}\xspace}
\newcommand{\fourthjetpt}{p\ensuremath{_{\textrm{T}}^{\textrm{Jet4}}}\xspace}
\newcommand{\leadbpt}{p\ensuremath{_{\textrm{T}}^{\textrm{b1}}}\xspace}
\newcommand{\leadleppt}{p\ensuremath{_{\textrm{T}}^{\textrm{l1}}}\xspace}
\newcommand{\eventsph}{\ensuremath{S}\xspace}
\newcommand{\hth}{\ensuremath{C}\xspace}
\newcommand{\drll}{dR\ensuremath{_{\textrm{ll}}}\xspace}
\newcommand{\drbb}{dR\ensuremath{_{\textrm{bb}}}\xspace}
\newcommand{\leadlepeta}{\ensuremath{\eta^{\textrm{l1}}}\xspace}
\newcommand{\redhadmass}{M\ensuremath{_{\textrm{RE}}^{\textrm{H}}}\xspace}
\newcommand{\sigmatttt}{\ensuremath{\sigma_{\tttt}}\xspace}
\newcommand{\ourLumi}{\ensuremath{2.6~{\rm fb}^{-1}}\xspace}
\newcommand{\invfb}{\ensuremath{\textrm{fb}^{-1}}\xspace}
\newcommand{\fb}{\ensuremath{\textrm{fb}}\xspace}
\newcommand{\sigmattttsm}{\ensuremath{\sigma_{\tttt}^{SM}}\xspace}
\newcommand{\BDTtrijetone}{\ensuremath{\textrm{BDT}_{\textrm{trijet1}}}\xspace}
\newcommand{\BDTtrijettwo}{\ensuremath{\textrm{BDT}_{\textrm{trijet2}}}\xspace}